\documentclass{article}
\usepackage{spconf,amsmath,graphicx,hyperref}
\usepackage[utf8]{inputenc} %
\usepackage[T1]{fontenc}    %
\usepackage{url}            %
\usepackage{booktabs}       %
\usepackage{amsfonts}       %
\usepackage{nicefrac}       %
\usepackage{microtype}      %
\usepackage{xcolor}         %
\usepackage{subcaption}    %
\usepackage{longtable}
\usepackage{multirow}
\usepackage[capitalize]{cleveref}
\usepackage{svg}
\usepackage[most]{tcolorbox}
\usepackage{wrapfig}
\usepackage{booktabs}

\title{Bias beyond Borders: Global Inequalities in AI-Generated Music}
\name{Ahmet Solak \qquad Florian Grötschla \qquad Luca A. Lanzendörfer \qquad Roger Wattenhofer}
\address{ETH Zurich}
\begin{document}
\maketitle
\begin{abstract}
While recent years have seen remarkable progress in music generation models, research on their biases across countries, languages, cultures, and musical genres remains underexplored. This gap is compounded by the lack of datasets and benchmarks that capture the global diversity of music. To address these challenges, we introduce GlobalDISCO, a large-scale dataset consisting of 73k music tracks generated by state-of-the-art commercial generative music models, along with paired links to 93k reference tracks in LAION-DISCO-12M. The dataset spans 147 languages and includes musical style prompts extracted from MusicBrainz and Wikipedia. The dataset is globally balanced, representing musical styles from artists across 79 countries and five continents. Our evaluation reveals large disparities in music quality and alignment with reference music between high-resource and low-resource regions. Furthermore, we find marked differences in model performance between mainstream and geographically niche genres, including cases where models generate music for regional genres that more closely align with the distribution of mainstream styles.
\end{abstract}
\section{INTRODUCTION}
\label{sec:intro}

\begin{figure}
  \centering
  \includegraphics[width=\columnwidth]{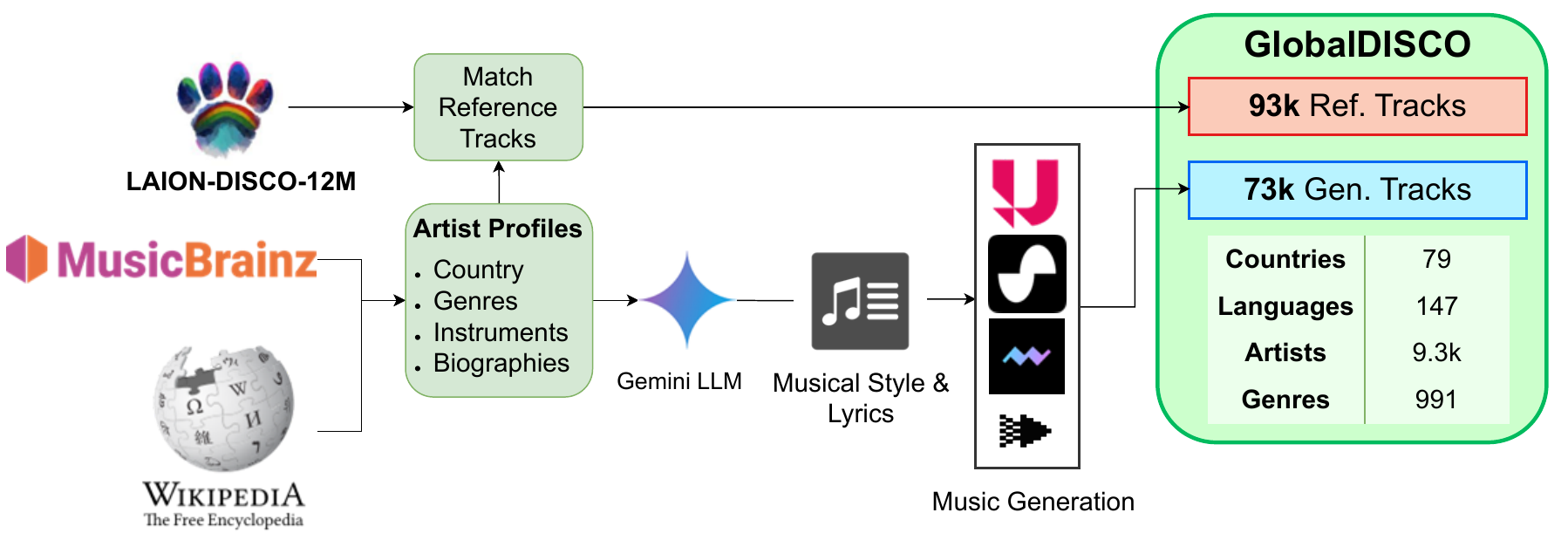}
  \caption{Pipeline of data collection and audio generation for GlobalDISCO. We gather artist information from MusicBrainz and Wikipedia, match it with reference tracks from LAION-DISCO-12M, and construct artist profiles based on this information. These profiles are then used to generate music using state-of-the-art music generation models, resulting in a globally diverse dataset of both generated tracks and reference tracks.}
  \label{fg:pipeline}
\end{figure}

\begin{figure}[h]
    \begin{center}
        \includegraphics[width=0.99\columnwidth]{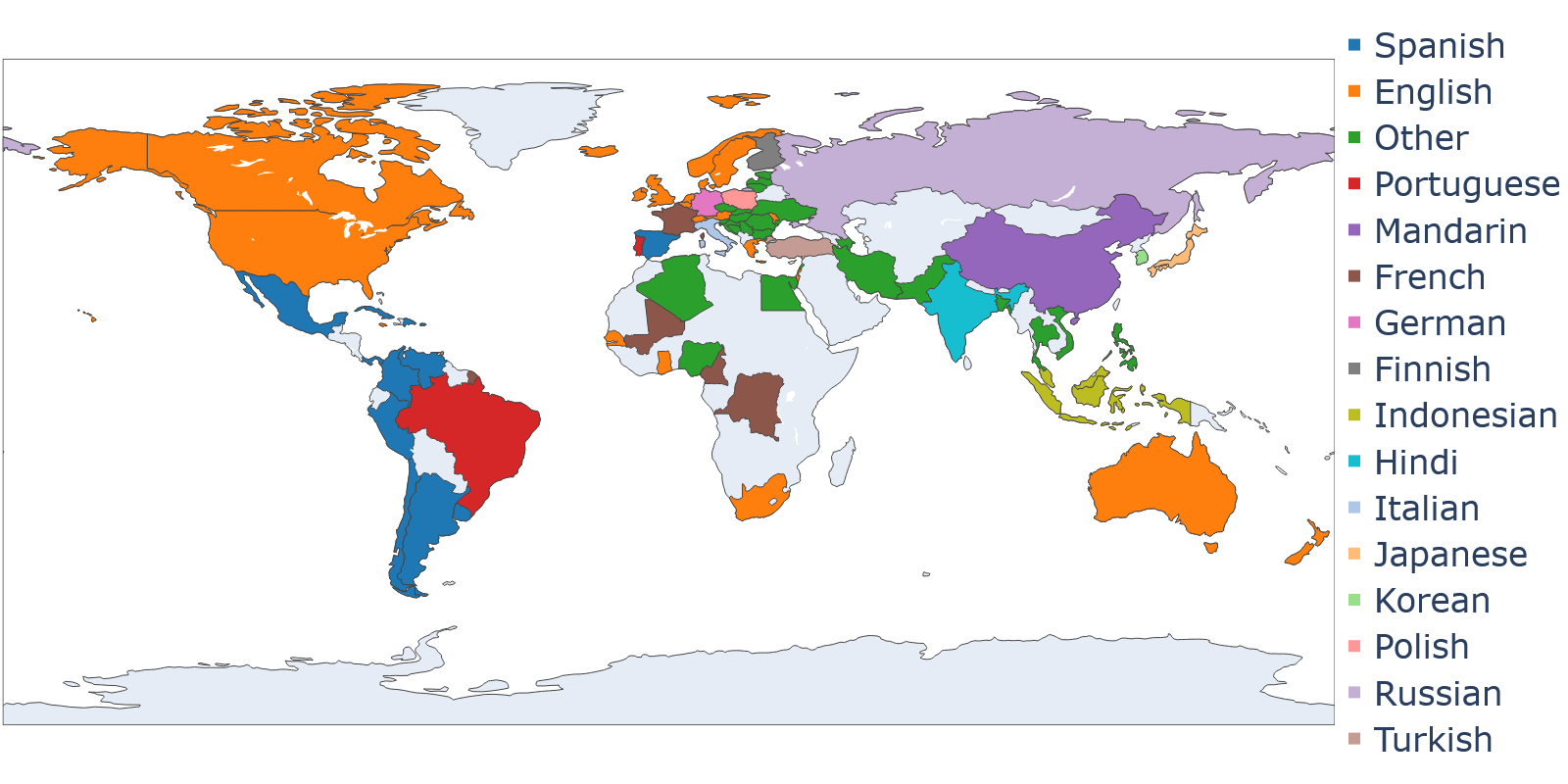}
        \caption{World map with all 79 countries represented in GlobalDISCO with the majority language of generated music denoted by color. Each country has a minimum of 75 generated tracks, with a median of 502 and a maximum of 2,861.}
        \label{fg:worldmap_countries}
    \end{center}
\end{figure}

In terms of quality and performance, the music generation field has seen remarkable progress in recent years, with commercial systems achieving exceptional results, even outperforming real music in large-scale human evaluation studies \cite{grotschla2025benchmarking}. However, despite music being a universal human experience found in all cultures around the world \cite{hodges2019music}, recent studies have highlighted a significant lack of intercultural and multilingual datasets in music generation research \cite{mehta2025musicallexploringmulticultural}. These findings, combined with the rapid progress of generative models, further underscore the urgent need for resources that allow the evaluation of potential biases and weaknesses in these models.
In other domains, biases across world regions and cultures have been more widely researched, with benchmarks and datasets released that aim to address intercultural biases in both the image \cite{kannen2025aestheticsculturalcompetencetexttoimage} and language domains \cite{myung2025blendbenchmarkllmseveryday,romero2024cvqaculturallydiversemultilingualvisual}.
To the best of our knowledge, the only publicly available multilingual generated music dataset~\cite{li2024m6multigeneratormultidomainmultilingual} contains only music in 3 different languages.
In comparison, the recently published BLEND \cite{myung2025blendbenchmarkllmseveryday} and CVQA \cite{romero2024cvqaculturallydiversemultilingualvisual} benchmarks for large language models and multimodal large language models have global coverage, with BLEnD covering 13 languages and 16 different countries and CVQA covering 31 languages and 30 different countries.

To address these challenges in the field of music generation, we present the GlobalDISCO dataset,\footnote{\url{https://huggingface.co/datasets/disco-eth/GlobalDISCO}} which is designed to evaluate biases and diversity in music generation. GlobalDISCO consists of 93k real and 73k generated music from 79 countries, across five continents, and 147 languages. The tracks in GlobalDISCO are generated with four state-of-the-art commercial models: Udio~\cite{udio}, Suno~\cite{suno}, Mureka~\cite{mureka}, and Riffusion~\cite{producerai}. Their performances and biases are explored across geographical regions and genres to provide a representative evaluation of the current capabilities and limitations of available music generation systems.

Analyzing GlobalDISCO, we find that state-of-the-art music generation models are highly biased across both world regions and genres, and that they generate music much more out-of-distribution for lower-resource regions and genres compared to higher-resource regions and mainstream genres. Furthermore, when instructed to generate music for certain regional genres, the models often produce music that is more closely aligned with the distribution of mainstream genres, both in terms of objective metrics and human perception.
By releasing GlobalDISCO as a public resource, we aim to support the research community in identifying and addressing biases in music generation and to promote greater global diversity in future model development.

\section{METHODOLOGY}
\label{sec:method}

\subsection{Dataset construction}
\label{ssec:dataset}
To construct the GlobalDISCO dataset, we begin by collecting artist entries from MusicBrainz, selecting those that include information about the artist's geographical area, as well as links to additional biographical pages. This initial step gives us 148k artist profiles. For artists without Wikipedia articles linked directly from their MusicBrainz pages, we perform supplementary searches using artist names on English Wikipedia. We select articles that match the MusicBrainz profiles by name and at least two additional attributes, such as area or genre tags. We enrich this artist metadata with reference tracks from the LAION-DISCO-12M \cite{lanzendörfer2023disco10mlargescalemusicdataset,laion2024laiondisco12m} dataset by matching artist and channel names, as well as verifying discography overlap when there is more than a single match. We retain the top-10 most viewed tracks per artist, with view numbers taken from LAION-DISCO-12M. To focus on music with vocals in different languages, we exclude artists associated only with instrumental genres. This approach retains artists with instrumental genres such as classical and electronic when their metadata also includes vocal genres.
From the 34k artists that fulfill this criterion, we ensure a balanced global representation by selecting up to a threshold \(t=374\) artists per country, where the value of \(t\) is determined via binary search to yield a dataset with roughly 10k artists. For each artist, we construct a profile using the collected metadata and generate musical style descriptions and synthetic lyrics for the artists using Gemini~\cite{geminiteam2025geminifamilyhighlycapable}. \cref{fig:artist-profile} shows a sample artist profile with different sections, such as genre and instruments, as well as relevant biographical information. The musical style description generated with that artist profile is: ``turkish folk music, sufi music, arabesque music, anatolian rock, pop/rock. Male vocals, vocal harmonies. Satirical lyrics, spiritual themes.''. For lyrics we adopt the methodology of previous works \cite{rahman2025sonicssyntheticidentifying} to use real lyrics and then generate synthetic lyrics from up to three real samples with few-shot inference \cite{labrak2025syntheticlyricsdetectionlanguages}. For artists without real samples, we use the artist profile to generate lyrics. 

\begin{figure}[t]
\centering
\begin{tcolorbox}[colback=gray!5!white, colframe=blue!50!white, title=Artist Profile]

\textbf{Artist Name:} [Artist Name]

\textbf{Country:} türkiye

\textbf{Genres:} pop; rock, Turkish folk music, Sufi music, Arabesque music, Anatolian Rock 

\textbf{Active Dates:} [Active Dates] 

\textbf{Biography Language:} English 

\textbf{Biography:}
[Artist Name] was a Turkish pop and rock band consisted of members [Members]. While many of their songs poke fun at common Turkish types or satirise prejudice and corruption ...

\end{tcolorbox}
\caption{An artist profile constructed with information gathered from MusicBrainz and Wikipedia. The artist’s name (in this case, a band), the names of its members, and the active dates are illustrated here with placeholders.}
\label{fig:artist-profile}
\end{figure}

Using these prompts and lyrics, we generate music with four state-of-the-art music generation models: Suno (v4)~\cite{suno}, Udio (v1.5 Allegro)~\cite{udio}, Mureka (v6)~\cite{mureka}, and Riffusion (FUZZ 0.8)~\cite{producerai}. All four models are commercial black boxes that take musical styles and lyrics as textual input to generate music tracks. The structure and length of the generated musical styles and lyrics are chosen to make them suitable for all four models.

The final dataset includes 9.3k artists, for which all models were successfully able to generate music and reference tracks were available in LAION-DISCO-12M. 79 countries across 5 continents are represented in the dataset, with a minimum of 10 artists per country. We identify more than 991 genres using the list of available genre tags on MusicBrainz as well as 147 different languages among our generated lyrics using the GlotLID language identification model \cite{kargaran2023glotlid}. 18 of those languages have more than 100 different artists associated with them. A world map showing all countries in the dataset, as well as the majority language among their generated tracks is presented in \cref{fg:worldmap_countries}. In \cref{tab:dataset_comparison}, we compare GlobalDISCO to other open-sourced synthetic music datasets across various metrics. GlobalDISCO is larger in scale and more diverse in terms of language coverage compared to previous work. It also includes music from four state-of-the-art generation platforms, the most of any generative music dataset.

A high-level overview of the entire data collection, data processing, and music generation pipeline for GlobalDISCO is shown in \cref{fg:pipeline}.

\begin{table}[t]
\centering
\footnotesize
\renewcommand{\arraystretch}{1.2}
\setlength{\tabcolsep}{2pt}
\begin{tabular}{lccccc}
\toprule
& 
\shortstack{\textbf{SONICS} \\ \cite{rahman2025sonicssyntheticidentifying}} & 
\shortstack{\textbf{M6} \\ \cite{li2024m6multigeneratormultidomainmultilingual}} & 
\shortstack{\textbf{FakeMusicCaps} \\ \cite{comanducci2024fakemusiccapsdatasetdetectionattribution}} & 
\shortstack{\textbf{AIME} \\ \cite{grotschla2025benchmarking}} & 
\shortstack{\textbf{GlobalDISCO} \\ (Ours)} \\ 
\midrule
Gen. Tracks     & 49,074 & 9,194 & 27,605 & 6,000 & \textbf{73,792} \\
Ref. Tracks     & 48,090 & 4,299 & 5,521 & 500 & \textbf{92,859} \\
Models          & 5      & 6     & 5      & \textbf{12} & 4 \\
Languages       & 1      & 3     & 0      & 0 & \textbf{147} \\
Lyrics          & \textbf{Yes} & No & No & No & \textbf{Yes} \\
\bottomrule
\end{tabular}
\caption{Comparison of synthetic music datasets across various metrics.}
\label{tab:dataset_comparison}
\end{table}

\begin{figure}
    \begin{center}
        \includegraphics[width=0.98\columnwidth]{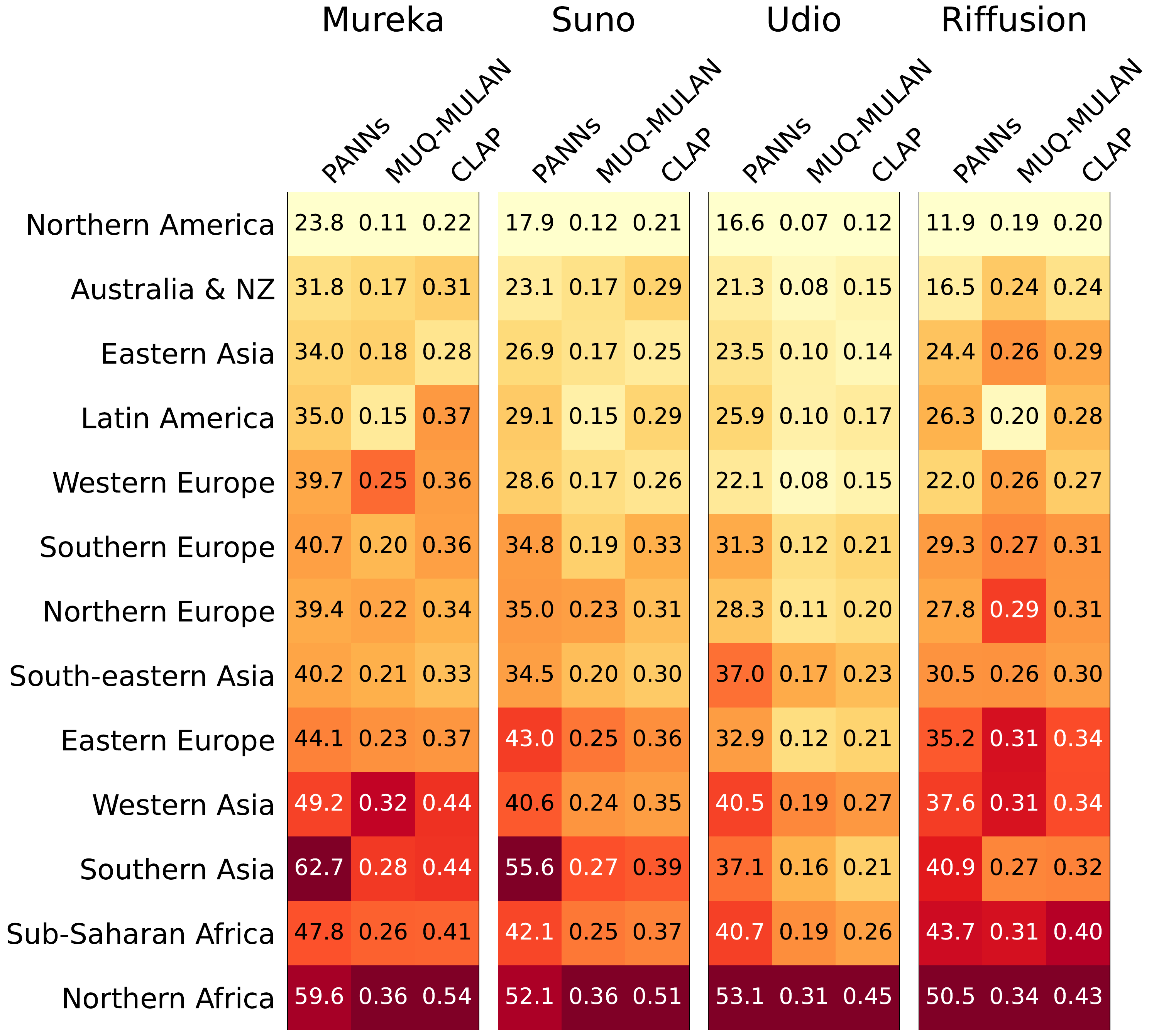}
        \caption{Mean FAD scores (lower is better), averaged across the countries for world regions. The regions are ordered by the mean z-scored FAD scores across embeddings. We find that the similarities of distributions between generated and reference tracks vary greatly between higher-resource regions (e.g., Northern America) and lower-resource regions (e.g., Sub-Saharan Africa).}
        \label{fg:region_fad}
    \end{center}
\end{figure}

\subsection{Evaluation}
\label{ssec:evaluation}
To evaluate generated and reference music, we use several audio embedding models. We use the PANNs \cite{kong2020panns} and CLAP \cite{laionclap2023} audio embedding models, which have shown good alignment with human preference in prior work \cite{grotschla2025benchmarking,gui2024adaptingfrechetaudiodistance}. For CLAP we choose the ``music\_audioset\_epoch\_15\_esc\_90.14'' checkpoint. As CLAP takes 10 second audio inputs, we compute embeddings for 10-second windows across the tracks with 1-second hops and then take the mean of those embeddings as the final CLAP embedding per track. We also select the MUQ-MULAN model \cite{zhu2025muqselfsupervisedmusicrepresentation}, which reports state-of-the-art results on music tagging tasks.
Using these embedding models, we use the Frechet Audio Distance (FAD) \cite{kilgour2019frechetaudiodistancemetric} and the Kernel Audio Distrance (KAD) \cite{chung2025kadfadeffectiveefficient} metrics. FAD compares evaluation and reference audio sets by comparing their multivariate Gaussian distributions. KAD is a more recently proposed distribution-free alternative, which is based on the Maximum Mean Discrepancy \cite{JMLR:v13:gretton12a}. For the kernel function in KAD we use the Gaussian radial basis function kernel, as proposed by the authors \cite{chung2025kadfadeffectiveefficient}. For both FAD and KAD lower scores are better.

\begin{figure}
    \begin{center}
        \includegraphics[width=0.98\columnwidth]{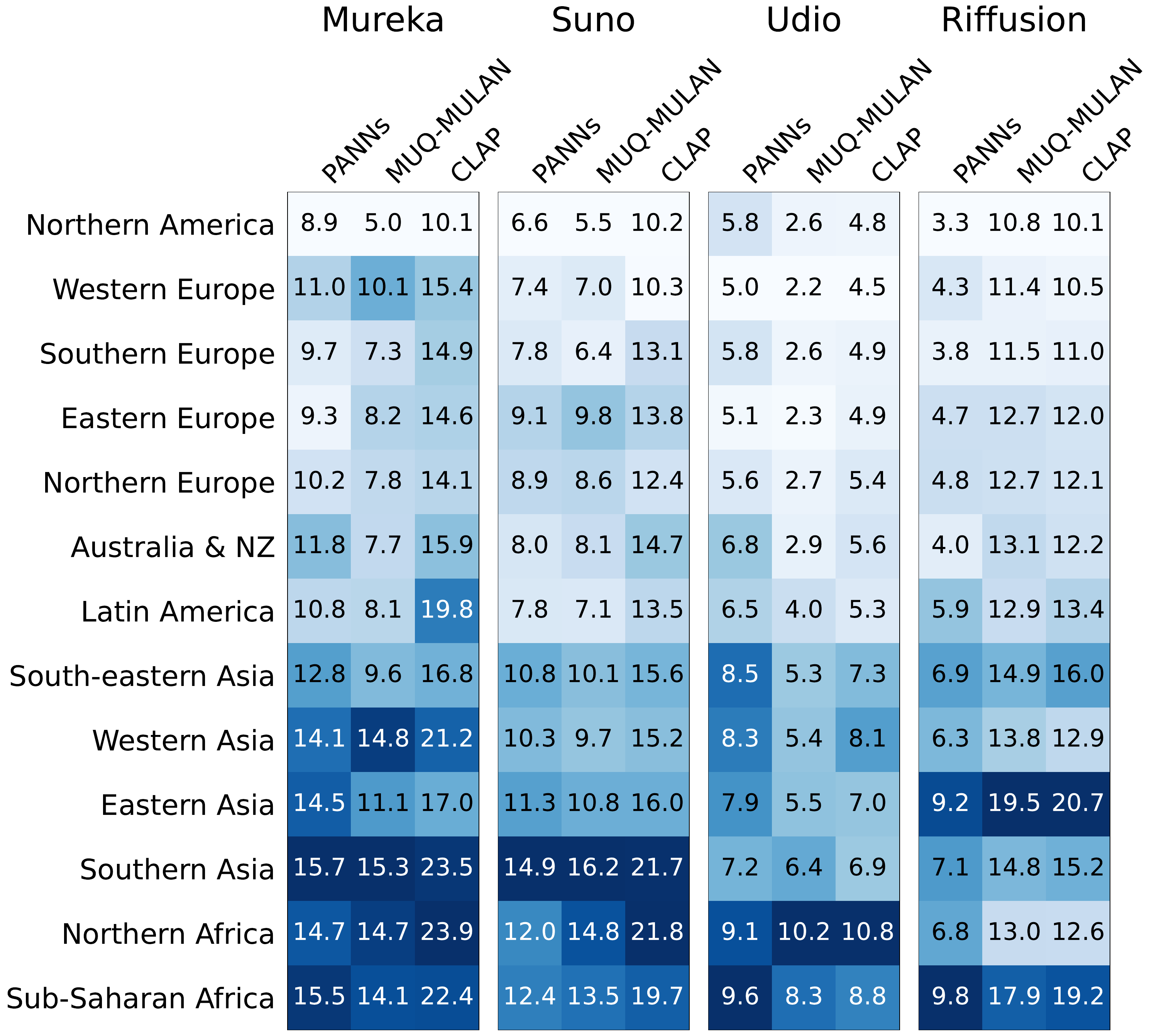}
        \caption{Mean KAD scores (lower is better), averaged across the countries for world regions. The regions are ordered by the mean z-scored KAD scores across embeddings. Similar to the FAD results, the similarities of distributions between generated and reference tracks vary greatly between higher- and lower resource regions.}
        \label{fg:region_kad}
    \end{center}
\end{figure}

\section{RESULTS}
\label{sec:results}

We first explore the difference in music generation quality across different world sub-regions, as defined by the UN M49 Standard~\cite{UNSD_M49_web_bibtex}. For the PANNs, CLAP, and MUQ-MULAN embedding models, we present FAD scores between generated and reference tracks for the 13 world regions present in GlobalDISCO, shown as a heatmap in \cref{fg:region_fad}. We show the same analysis with KAD in \cref{fg:region_kad}. The results indicate that model performance varies significantly across higher- and lower-resource regions. For all music generation models, world regions from the continent of Africa, as well as Southern and Western Asia, generate music that is considerably more out-of-distribution compared to higher-resource regions. At the other end, Northern America, which is likely the highest-resource region in terms of available music, shows better results across the board. In general, these trends are consistent between the embedding models and evaluation metrics, indicating the reliability and generality of the findings.

\begin{figure}
  \centering
  \includegraphics[width=0.98\columnwidth]{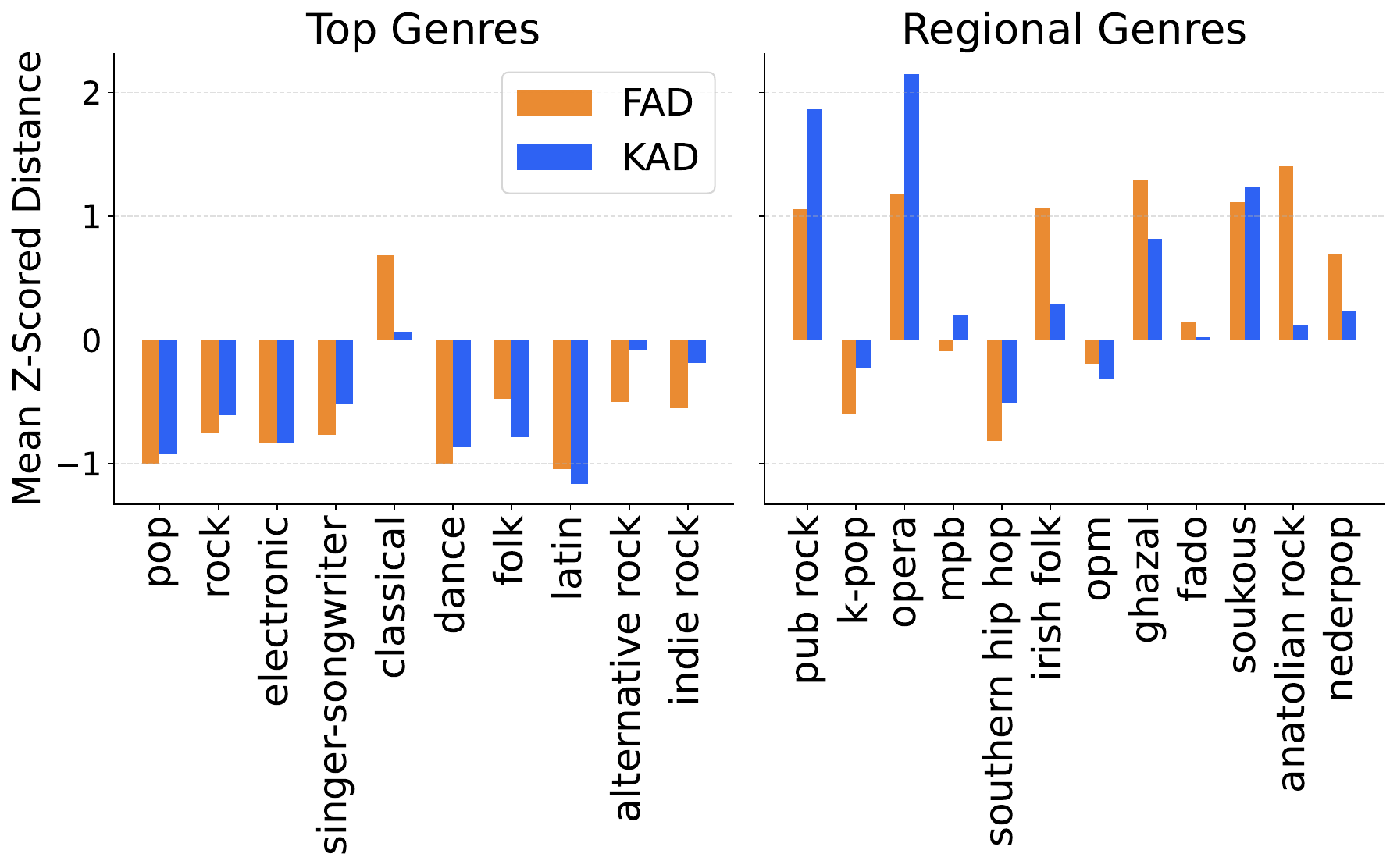}
  \caption{Mean normalized (z-scored) FAD and KAD scores across the top ten most popular (left) and top regional music genres (right). Scores are first normalized across genres for each combination of music generation and embedding model, and then averaged across all such combinations. We see that state-of-the-art models exhibit worse FAD and KAD scores for regional genres compared to mainstream genres.}
  \label{fg:mean_genre_bars}
\end{figure}

In \cref{fg:mean_genre_bars}, we show the mean normalized (z-scored) FAD and KAD scores for popular genres and regional genres, averaged across all embedding and music generation models. The ten most popular genres are selected by frequency in the dataset, whereas the regional genres are selected using a tf-idf-like method, which we compute by multiplying the relative frequency of a genre within a country by the inverse of the number of countries containing that genre.
For each region, we select the genre with the highest tf-idf-like score, provided that it is associated with at least 10 artists in the respective country. For Eastern Europe, opera is selected since the top two scoring genres, pop and classical, are already included among the popular genres.
Certain regional styles which have a wide global listening audience and great amounts of easily accessible audio online, such as southern hip hop and k-pop, exhibit FAD and KAD scores close to those of top mainstream styles.
However, most regional genres score significantly worse than mainstream music. Challenging regional genres include music from lower-resource regions, such as soukous from Sub-Saharan Africa, but also from higher-resource regions, such as pub rock from Australia. We also find that state-of-the-art models struggle more with generating in-distribution audio for traditional genres like opera and classical, compared to more modern styles.

\begin{figure}
  \centering
  \includegraphics[width=0.98\columnwidth]{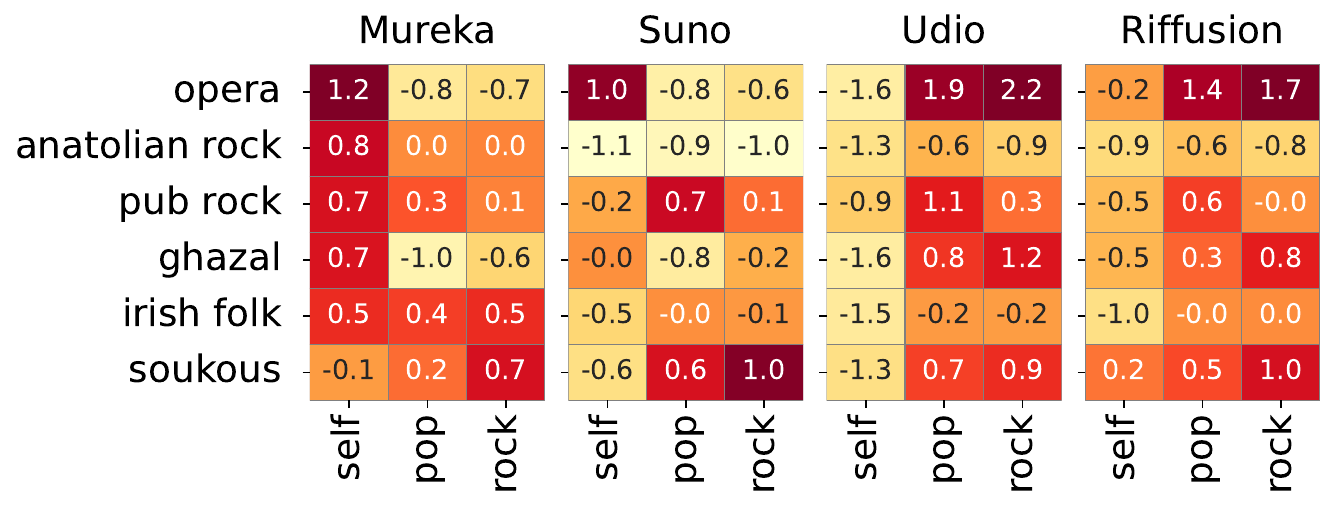}
  \caption{Normalized (z-scored) distances (lower is better) between the six worst scoring regional genres, their own reference tracks, and the two mainstream genres pop and rock. FAD and KAD scores are normalized per embedding model, and we report the mean across both metrics and models.}
  \label{fg:regional_comparisons}
\end{figure}

Furthermore, we compare the distributions of the six regional genres with the worst scores across metrics with their reference tracks, as well as reference music for pop and rock, the two most frequent genres in the dataset. The results in \cref{fg:regional_comparisons} show that the Mureka and Suno models display a strong bias towards mainstream genres. Mureka generates music for five of the six selected regional genres that more closely resembles pop and rock than the corresponding reference tracks. Suno shows similar biases, with generated opera and ghazal music that are closer in distribution to real pop music than to the corresponding reference tracks of the same genre.

In addition to these objective results, we further demonstrate how these biases are also clear to human listeners. We select generated tracks for regional genres across models and identify their closest neighbors among mainstream genres. Cosine distances are computed across embedding models, and the closest neighbor for each generated track is determined by summing its distance rankings across models and selecting the track with the lowest aggregated ranking. The resulting examples are publicly available,\footnote{\url{https://a-b-solak.github.io/globaldisco/}} and were chosen as cases where the generated tracks were stylistically closer to their nearest mainstream neighbors than to their reference artist’s style, as confirmed by human listeners.

\section{CONCLUSION}
\label{sec:conclusion}

In this work, we presented GlobalDISCO, a large-scale generated music dataset encompassing musical traditions from around the world aimed at exploring the potential biases in music generation models and addressing the lack of large, multicultural, and multilingual datasets in the generative music domain. Our findings reveal substantial disparities in the ability of models to generate music from low-resource regions, such as Northern Africa, Sub-Saharan Africa, and Southern Asia. We also observe genre-specific biases, where models not only have difficulty generating music for regional genres, but also generate audio for some of those genres that aligns more closely with mainstream genres such as pop and rock. As generated music continues to grow in popularity and quality, our results highlight clear biases against lower-resource musical traditions and the need to address them to preserve global musical diversity.

\vfill\pagebreak
\ninept
\bibliographystyle{IEEEbib}
\bibliography{strings,refs}

\begin{thebibliography}{10}

\bibitem{grotschla2025benchmarking}
Florian Gr{\"o}tschla, Ahmet Solak, Luca~A Lanzend{\"o}rfer, and Roger Wattenhofer,
\newblock ``Benchmarking music generation models and metrics via human preference studies,''
\newblock in {\em ICASSP 2025-2025 IEEE International Conference on Acoustics, Speech and Signal Processing (ICASSP)}. IEEE, 2025, pp. 1--5.

\bibitem{hodges2019music}
Donald~A Hodges,
\newblock {\em Music in the human experience: An introduction to music psychology},
\newblock Routledge, 2019.

\bibitem{mehta2025musicallexploringmulticultural}
Atharva Mehta, Shivam Chauhan, Amirbek Djanibekov, Atharva Kulkarni, Gus Xia, and Monojit Choudhury,
\newblock ``Music for all: Exploring multicultural representations in music generation models,'' 2025.

\bibitem{kannen2025aestheticsculturalcompetencetexttoimage}
Nithish Kannen, Arif Ahmad, Marco Andreetto, Vinodkumar Prabhakaran, Utsav Prabhu, Adji~Bousso Dieng, Pushpak Bhattacharyya, and Shachi Dave,
\newblock ``Beyond aesthetics: Cultural competence in text-to-image models,'' 2025.

\bibitem{myung2025blendbenchmarkllmseveryday}
Junho Myung, Nayeon Lee, Yi~Zhou, Jiho Jin, Rifki~Afina Putri, Dimosthenis Antypas, Hsuvas Borkakoty, Eunsu Kim, Carla Perez-Almendros, Abinew~Ali Ayele, Víctor Gutiérrez-Basulto, Yazmín Ibáñez-García, Hwaran Lee, Shamsuddeen~Hassan Muhammad, Kiwoong Park, Anar~Sabuhi Rzayev, Nina White, Seid~Muhie Yimam, Mohammad~Taher Pilehvar, Nedjma Ousidhoum, Jose Camacho-Collados, and Alice Oh,
\newblock ``Blend: A benchmark for llms on everyday knowledge in diverse cultures and languages,'' 2025.

\bibitem{romero2024cvqaculturallydiversemultilingualvisual}
David Romero et~al.,
\newblock ``Cvqa: Culturally-diverse multilingual visual question answering benchmark,'' 2024.

\bibitem{li2024m6multigeneratormultidomainmultilingual}
Yupei Li, Hanqian Li, Lucia Specia, and Björn~W. Schuller,
\newblock ``M6: Multi-generator, multi-domain, multi-lingual and cultural, multi-genres, multi-instrument machine-generated music detection databases,'' 2024.

\bibitem{udio}
Udio,
\newblock ``Udio,'' \url{https://www.udio.com/}, 2025,
\newblock Accessed: April-May, 2025.

\bibitem{suno}
Suno,
\newblock ``Suno,'' \url{https://suno.com/}, 2025,
\newblock Accessed: April-May, 2025.

\bibitem{mureka}
Mureka,
\newblock ``Mureka,'' \url{https://www.mureka.ai/}, 2025,
\newblock Accessed: April-May, 2025.

\bibitem{producerai}
Producer.ai,
\newblock ``Producer.ai,'' \url{https://www.producer.ai/}, 2025,
\newblock Accessed: April-August, 2025.

\bibitem{lanzendörfer2023disco10mlargescalemusicdataset}
Luca~A. Lanzendörfer, Florian Grötschla, Emil Funke, and Roger Wattenhofer,
\newblock ``Disco-10m: A large-scale music dataset,'' 2023.

\bibitem{laion2024laiondisco12m}
{LAION e.V.},
\newblock ``Laion-disco-12m,'' \url{https://huggingface.co/datasets/laion/LAION-DISCO-12M}, 2024,
\newblock Dataset of 12 million YouTube music links with metadata; Apache‑2.0 license.

\bibitem{geminiteam2025geminifamilyhighlycapable}
Gemini Team et~al.,
\newblock ``Gemini: A family of highly capable multimodal models,'' 2025.

\bibitem{rahman2025sonicssyntheticidentifying}
Md~Awsafur Rahman, Zaber Ibn~Abdul Hakim, Najibul~Haque Sarker, Bishmoy Paul, and Shaikh~Anowarul Fattah,
\newblock ``Sonics: Synthetic or not -- identifying counterfeit songs,'' 2025.

\bibitem{labrak2025syntheticlyricsdetectionlanguages}
Yanis Labrak, Markus Frohmann, Gabriel Meseguer-Brocal, and Elena~V. Epure,
\newblock ``Synthetic lyrics detection across languages and genres,'' 2025.

\bibitem{kargaran2023glotlid}
Amir~Hossein Kargaran, Ayyoob Imani, Fran{\c{c}}ois Yvon, and Hinrich Sch{\"u}tze,
\newblock ``{G}lot{LID}: Language identification for low-resource languages,''
\newblock in {\em The 2023 Conference on Empirical Methods in Natural Language Processing}, 2023.

\bibitem{comanducci2024fakemusiccapsdatasetdetectionattribution}
Luca Comanducci, Paolo Bestagini, and Stefano Tubaro,
\newblock ``Fakemusiccaps: a dataset for detection and attribution of synthetic music generated via text-to-music models,'' 2024.

\bibitem{kong2020panns}
Qiuqiang Kong, Yin Cao, Turab Iqbal, Yuxuan Wang, Wenwu Wang, and Mark~D Plumbley,
\newblock ``Panns: Large-scale pretrained audio neural networks for audio pattern recognition,''
\newblock {\em IEEE/ACM Transactions on Audio, Speech, and Language Processing}, vol. 28, pp. 2880--2894, 2020.

\bibitem{laionclap2023}
Yusong Wu*, Ke~Chen*, Tianyu Zhang*, Yuchen Hui*, Taylor Berg-Kirkpatrick, and Shlomo Dubnov,
\newblock ``Large-scale contrastive language-audio pretraining with feature fusion and keyword-to-caption augmentation,''
\newblock in {\em IEEE International Conference on Acoustics, Speech and Signal Processing, ICASSP}, 2023.

\bibitem{gui2024adaptingfrechetaudiodistance}
Azalea Gui, Hannes Gamper, Sebastian Braun, and Dimitra Emmanouilidou,
\newblock ``Adapting frechet audio distance for generative music evaluation,'' 2024.

\bibitem{zhu2025muqselfsupervisedmusicrepresentation}
Haina Zhu, Yizhi Zhou, Hangting Chen, Jianwei Yu, Ziyang Ma, Rongzhi Gu, Yi~Luo, Wei Tan, and Xie Chen,
\newblock ``Muq: Self-supervised music representation learning with mel residual vector quantization,'' 2025.

\bibitem{kilgour2019frechetaudiodistancemetric}
Kevin Kilgour, Mauricio Zuluaga, Dominik Roblek, and Matthew Sharifi,
\newblock ``Fr\'echet audio distance: A metric for evaluating music enhancement algorithms,'' 2019.

\bibitem{chung2025kadfadeffectiveefficient}
Yoonjin Chung, Pilsun Eu, Junwon Lee, Keunwoo Choi, Juhan Nam, and Ben~Sangbae Chon,
\newblock ``Kad: No more fad! an effective and efficient evaluation metric for audio generation,'' 2025.

\bibitem{JMLR:v13:gretton12a}
Arthur Gretton, Karsten~M. Borgwardt, Malte~J. Rasch, Bernhard Sch{{\"o}}lkopf, and Alexander Smola,
\newblock ``A kernel two-sample test,''
\newblock {\em Journal of Machine Learning Research}, vol. 13, no. 25, pp. 723--773, 2012.

\bibitem{UNSD_M49_web_bibtex}
{United Nations Statistics Division},
\newblock ``Standard country or area codes for statistical use (m49),'' \url{https://unstats.un.org/unsd/methodology/m49/},
\newblock Accessed 27 Aug 2025. Online version of United Nations publication Series M, No.~49.

\end{thebibliography}

\end{document}